\documentclass[a4paper]{jpconf}
\usepackage[utf8]{inputenc}
\usepackage{amsmath}
\usepackage{hyperref}
\usepackage{braket}
\usepackage{tikz}
\usetikzlibrary{arrows, positioning, automata}
\begin{document}
\title{Scattering amplitude and its soft decomposition}

\author{Andriniaina Narindra Rasoanaivo}

\address{Sciences Expérimentales et des Mathématiques, 
Ecole Normale Supérieure d'Antananarivo, \\ Complexe Scolaire Ampefiloha BP 881, Université d'Antananarivo, Madagascar}

\ead{andriniaina@aims.ac.za}

\begin{abstract}
In the pure scattering theory, the universality of the soft limit has been studied for a long time. In this talk we review the property of soft limit to relate an $n$-point amplitude to an $(n-1)$-point amplitude. We show how this property can be used to decompose amplitudes into different complementary soft channel. The existence of such decomposition provides a new way to understand how to construct amplitude solely from them soft limit.
\end{abstract}

\section{Introduction}
Many progress has been made to unravel the mathematical structure of scattering amplitudes in gauge theories and gravity. These mathematical structure are often hidden in the complexity of the traditional  Feynman diagram expansion computation. 
During the past few decades, the development of modern techniques was so popular to make the different mathematical structure of amplitudes manifests: Britto Cachazo Fen Witten (BCFW) recursion, MHV rules, generalized unitarity cut, \ldots The most popular among these approaches is the BCFW recursion relations, they were developed allowing any tree level helicity amplitude to be calculated using a simple on-shell recursion relation. The on-shell recursion is used recursively to link higher points amplitudes to the lower points, in which three points are the seeds of the recursion \cite{Britto:2004ap}. Central to this construction have been unitarity, locality and the Wigner's little group which is the subgroup of the Lorentz group that leaves a momentum invariant. 

In this work, we will focus our attention on the factorization structure of amplitudes in the case where one gluon is going soft, i.e. the energy of the single gluon approaches zero. Such factorization is well understood from the Weinberg's theorem in which the physical properties of the low energy contribution are factorized out from the main amplitudes \cite{Weinberg:1965nx, Elvang:2016qvq, Jackiw:1968zza}. The soft contribution is known to be an operator acting on the lower point amplitude emerges from the factorization such that 

%
\begin{equation}\label{weinberg}
A_n(p_1,\ldots, p_n)
\xrightarrow{\quad p_n\to 0\quad}
\hat{S}(p_n)\times A_{n-1}(p_1,\ldots,p_{n-1}), 
\end{equation}
where $A_n$ is the main amplitude, $A_{n-1}$ is the emerged lower amplitude and $\hat{S}(p_n)$ is the soft operator associated to the soft particle of momentum $p_n$. 

This soft factorization is an universal property of amplitudes in which the soft operator is proven to be computable directly from the asymptotic symmetries of the associated soft particle \cite{Rasoanaivo:2020yii, Campoleoni:2017mbt}. The main objective of this work is to investigate the possibility of constructing any tree-level helicity amplitude of gluons from them asymptotic symmetries by inverting the soft factorization. Such approach is also known as the invert soft limit (ISL) which is a well defined construction for super Young Mills theory \cite{Nandan:2012rk}, however in \cite{Boucher-Veronneau:2011rwd} they showed that for the case of gluon amplitudes the inverse soft limit construction constructed form BCFW recursion is only possible up to 7-points for any given helicity configuration. 

We expect that if we understand the underline structure of amplitudes, which makes the soft factorization manifest, we might able to reconstruct any given amplitude from them asymptotic symmetries. To achieve that goal we will present in this work how a given amplitude can be decomposed in two pieces around one of its soft factorization; and we will use this decomposition to show the possibility to compute higher point amplitude from them lower points through inverse soft limits. In the section 2 we will give a short review on Weinberg's theorem and soft operators, then in section 3 we will present the idea of soft decomposition of a given amplitude, and we will conclude this work in section 4.

\section{Weinberg's theorem and the helicity soft operators}

For simplicity, we will use spinor variables to describe Weinberg's theorem. It is important to mention that this choice of variables will not change our description due to the universality of the theorem \cite{Weinberg:1965nx,Higuchi:2018vyu}. Therefore, through this work, we consider scattering amplitudes of gluons with external state labeled by $i=\{\lambda_i^a,\bar{\lambda}_i^{\dot{a}}\}$ and $h_i$. The variables $\lambda$ and $\bar{\lambda}$ are respectively a left and a right handed spinors such that $p_i^\mu (\sigma_\mu)^{a\dot{a}}=\lambda^a_i\bar{\lambda}^{\dot{a}}_i$, and $h_i$ is the helicity of the $i$-th external state which is either equal to $``+1"$ or $``-1"$, review on spinor variables \cite{Cheung:2017pzi,Britto:2005fq,Arkani-Hamed:2017jhn}. With these spinor variables, we can introduce two new invariant products
\begin{equation}
\braket{ij}= \varepsilon_{ab}\lambda_i^a\lambda^b_j \quad \text{and} \quad [ij]=\varepsilon_{\dot{a}\dot{b}}\bar{\lambda}_i^{\dot{a}}\bar{\lambda}_j^{\dot{b}}\; ,
\end{equation}
where $\varepsilon_{ab}$ and $\varepsilon_{\dot{a}\dot{b}}$ are the Kronecker symbols contracted with the spinor variables. 

The Weinberg's soft theorem is an universal property of scattering amplitudes, in which the soft contribution of the amplitude is factorize out from the high energy contribution through the relation \eqref{weinberg}. The soft contribution, denoted by $\hat{S}(p_s)$, is known to be an operator acting on the hard contribution the lower point amplitude emerged from the factorization. The soft operator is know to be a composition of the terms: \emph{the Weinberg soft factor} known as the leading term from the perturbative computation of $\hat{S}$, and \emph{a translation operator} represented as $\hat{U}$ which shifts the two adjacent momenta to the soft particle in the emergent lower point amplitude

\begin{equation}\label{soff_operator}
 \hat{S}(p_s)=S^{(0)}(p_s)\times \hat{U}(\delta_i,\delta_j ).
 \end{equation} 
In this equation, $i$ and $j$ are the labels of the momenta adjacent to the soft particle of momentum $p_s$, and $\delta_i$ and $\delta_j$ are the parameters of the translation such that $p_i\to p_i+\delta_i$ and $p_j\to p_j+\delta_j$. Depending  on the helicity $h_s$ of the soft momenta the soft operator and the parameters of the shifts are given by \cite{Boucher-Veronneau:2011rwd}
\begin{itemize}
\item[$\bullet$] for $h_s=+1$ we have 
\begin{equation}\label{positive_shifts}
S^{(0)}(p_s^+)= \frac{\braket{ij}}{\braket{is}\braket{sj}} 
\quad\text{and}\quad 
\left \{ \begin{aligned} 
\delta_i= \frac{\braket{js}}{\braket{ji}}\lambda_i\bar{\lambda}_s
\\
\delta_j= \frac{\braket{is}}{\braket{ij}}\lambda_j\bar{\lambda}_s
\end{aligned}\right .
\end{equation}
\item [$\bullet$] for $h_s=-1$ we have 
\begin{equation}\label{negative_shifts}
S^{(0)}(p_s^-)= \frac{[ij]}{[is][sj]} 
\quad\text{and}\quad 
\left \{ \begin{aligned} 
\delta_i= \frac{[js]}{[ji]}\lambda_s\bar{\lambda}_i
\\
\delta_j= \frac{[is]}{[ij]}\lambda_s\bar{\lambda}_j
\end{aligned}\right .
\end{equation}
\end{itemize}
here the momenta $p_i$, $p_j$ and $p_s$ are represented by them respective spinor variables $\lambda_i\bar{\lambda}_i$, $\lambda_j\bar{\lambda}_j$ and $\lambda_s\bar{\lambda}_s$. 
The Weinberg soft theorem is therefore a map from higher point amplitude to a lower points by factorizing the associated soft operator through the relation \eqref{weinberg}. From the description of the soft operator, applying the soft operator $\hat{S}$ on a lower point amplitude seems to invert the factorization and leads to the higher points which is not the case as we will discus in the following section.

\section{Soft decomposition}

\subsection{Soft structure of amplitudes}
In order to introduce the idea of soft decomposition of an amplitude, let us consider the six-point helicity amplitude of gluons in \eqref{amplitude}. With the six-point amplitude, we are going to investigate what happen when we perform the following limits: first we take one of the particle soft then, as a second step, we inverse the soft limit as in \cite{Boucher-Veronneau:2011rwd} using the soft operators defined from the previews section.

The derivation of the six-point gluon amplitude can be found in \cite{Britto:2004ap} which is given by

\begin{equation}\label{amplitude}
A_6(1^-,2^-,3^-,4^+,5^+,6^+) = \frac{1}{[2|3+4|5\rangle}\left (\frac{[4|5+6|1\rangle^3}{[34][23]\braket{56}\braket{61}S_{234}}+\frac{[6|1+2|3\rangle^3}{[61][12]\braket{34}\braket{45}S_{612}}\right ),
\end{equation}
where the other invariant quantities $S_{ijk}$ and $[i|j+k|l\rangle$ are defined as 
\begin{equation}
S_{ijk}=(p_i+p_j+p_k)^2 \quad\text{and} \quad [i|j+k|l\rangle=[ij]\braket{jl}+[ik]\braket{kl}.
\end{equation}
In this configuration of helicity, we can take any of six momenta to be soft. In our case, let us focus on the sixth momentum with which we will successively perform a soft then then an inverse soft limits. We know that taking a soft limit will factorize the soft behavior from the high energy regime, this is nothing than a map from the six-point to a five point amplitude 
\begin{equation}
A_6(1^-,2^-,3^-,4^+,5^+,6^+) \xrightarrow{\quad \text{soft } S(6^+)\quad }A_5(1^-,2^-,3^-,4^+,5^+).
\end{equation}
The $A_5(1^-,2^-,3^-,4^+,5^+)$ is known to be a MHV amplitude and its expression is given by the Park-Taylor formula \cite{Cheung:2017pzi}. In the expression \eqref{amplitude}, when we take sixth particle to go soft and as the momentum $p_6$ goes to zero, the second term tend to zero while the first term tend to factorize as Weinberg described in the theorem, i.e. only the first part contributes in the soft factorization of the sixth particle. 

Now we can invert the soft limit on the five-point amplitude in which we act the soft operator defined in \eqref{soff_operator} on the amplitude $A_5(1^-,2^-,3^-,4^+,5^+)$. That means we multiply the amplitude with the Weinberg soft factor $S^{(0)}(6^+)$ and shifts the momenta $p_5$ and $p_1$ according to the relation \eqref{positive_shifts}
\begin{equation}
A_5(1^-,2^-,3^-,4^+,5^+)\xrightarrow{\quad \text{inverse soft }  S(6^+) \quad}\frac{1}{[2|3+4|5\rangle}\times\frac{[4|5+6|1\rangle^3}{[34][23]\braket{56}\braket{61}S_{234}}.
\end{equation}
This above relation shows that the inverse soft limit of the sixth particle doesn't give us the original amplitude, however it only gives the first term in the expression of six-point amplitude back. The information on the term that tend to zero in soft limit was lost, and it can not be restored by inverting the soft limit. So viewed from the soft/inverse soft limit of the sixth particle the amplitude can be decomposed as follow
\begin{equation}
  A_6(1^-,2^-,3^-,4^+,5^+,6^+)= A_6^{[6]}(1^-,2^-,3^-,4^+,5^+,6^+)+R_6^{[6]}(1^-,2^-,3^-,4^+,5^+,6^+).
  \end{equation}  
  The $A_6^{[6]}$ is the central piece of the amplitude from the sixth particle's soft limit, and $R_6^{[6]}$ is the negligible term that tends to zero as the six goes soft,
   
  \begin{equation}
 \left \{ \begin{aligned}
&A_6^{[6]}(1^-,2^-,3^-,4^+,5^+,6^+) \xrightarrow{\quad \text{soft } S(6^+)\quad }A_5(1^-,2^-,3^-,4^+,5^+),\\
&R_6^{[6]}(1^-,2^-,3^-,4^+,5^+,6^+) \xrightarrow{\quad \text{soft } S(6^+)\quad } 0.
\end{aligned}\right .
  \end{equation}
Similar decomposition can be done with any of the momentum in amplitude. In a most general case, viewed from the softness of any $i$-th particle of a given $n$-point amplitude, the amplitude can be decomposed as
\begin{equation}\label{soft_decomposition}
A_n(1,2,\ldots,n)=A^{[i]}_n(1,2,\ldots,n)+R^{[i]}_n(1,2,\ldots,n), 
\end{equation}
where
\begin{itemize}
\item[$\bullet$] $A_n^{[i]}$ is the core of the soft theorem: which is central to the soft theorem as the $i$-th particle get soft that contains the IR behavior of the soft limit,
\item[$\bullet$] and $R_n^{[i]}$ is the soft shell of the amplitude: which is the part of the amplitude that is lost in the soft limit of the $i$-th particle.
\end{itemize}
This decomposition is similar to the decomposition presented in \cite{Rodina:2018pcb} but the reconstruction in the following section differs from them soft reconstruction of amplitudes.

\subsection{Soft reconstruction of amplitudes}

The soft decomposition \eqref{soft_decomposition} is not an unique decomposition, in fact for a given $n$-point amplitude there are $n$ different decomposition respectively to the $n$ particles. The soft decomposition also make the soft limit of a given particle manifests in which the soft shell $R_n^{[i]}$ will not contribute in the universal factorization of the amplitude. Since the core of the soft theorem $A_n^{[i]}$ is the only part that can be constructed form inverse soft limit, it is necessary to find an algorithm to reconstruct the soft shell. The study of the six-point amplitude \eqref{amplitude} can point us to the right direction for reconstruct $R_n^{[i]}$, for that we computed all the six different decomposition of the amplitude and found that the soft shell from one decomposition can be found in the central core of the other decomposition.

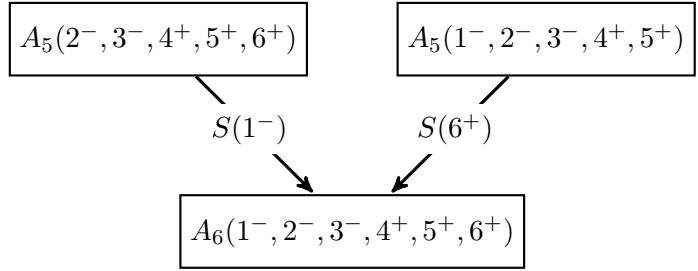
\begin{figure}
\centering
\usetikzlibrary{arrows, positioning, automata}
\begin{tikzpicture}[scale=.85,>=stealth',shorten >=1pt,node distance=2.5cm,on grid,initial/.style={rectangle},thick]
\def\p{*3};
\tikzset{mystyle/.style={->,very thick}}
\tikzset{every node/.style={fill=white}}
\node[state,initial] (T0) {$A_6(1^-,2^-,3^-,4^+,5^+,6^+)$};
\node[state,initial] (T1) at (1\p, 1\p) {$A_5(1^-,2^-,3^-,4^+,5^+)$};
\node[state,initial] (T2) at (-1\p, 1\p) {$A_5(2^-,3^-,4^+,5^+,6^+)$};

 \path (T1) edge [mystyle] node[pos=0.45] { $S(6^+)$} (T0);
  \path (T2) edge [mystyle] node[pos=0.45] { $S(1^-)$} (T0);

\end{tikzpicture}
\caption{Representation of the soft reconstruction of $A_6(1^-,2^-,3^-,4^+,5^+,6^+)$ }
\label{diagram_6}
\end{figure}

In order to see how this works, here are the six decomposition of the six-point amplitude 
\begin{equation}
\begin{aligned}
A_6(1^-,2^-,3^-,4^+,5^+,6^+)&= A^{[1]}_6(1^-,2^-,3^-,4^+,5^+,6^+)+ R^{[1]}_6(1^-,2^-,3^-,4^+,5^+,6^+)\\
&= A^{[2]}_6(1^-,2^-,3^-,4^+,5^+,6^+)+ R^{[2]}_6(1^-,2^-,3^-,4^+,5^+,6^+)\\
&= A^{[3]}_6(1^-,2^-,3^-,4^+,5^+,6^+)+ R^{[3]}_6(1^-,2^-,3^-,4^+,5^+,6^+)\\
&= A^{[4]}_6(1^-,2^-,3^-,4^+,5^+,6^+)+ R^{[4]}_6(1^-,2^-,3^-,4^+,5^+,6^+)\\
&= A^{[5]}_6(1^-,2^-,3^-,4^+,5^+,6^+)+ R^{[5]}_6(1^-,2^-,3^-,4^+,5^+,6^+)\\
&= A^{[6]}_6(1^-,2^-,3^-,4^+,5^+,6^+)+ R^{[6]}_6(1^-,2^-,3^-,4^+,5^+,6^+)
\end{aligned}
\end{equation}
in which we have the following expressions
\begin{equation}
\left \{\begin{aligned}
&A_6^{[1]}=A_6^{[4]}=R_6^{[3]}=R_6^{[6]}=\frac{1}{[2|3+4|5\rangle}\left (\frac{[6|1+2|3\rangle^3}{[61][12]\braket{34}\braket{45}S_{612}}\right ),\\
&A_6^{[3]}=A_6^{[6]}=R_6^{[1]}=R_6^{[4]}=\frac{1}{[2|3+4|5\rangle}\left (\frac{[4|5+6|1\rangle^3}{[34][23]\braket{56}\braket{61}S_{234}}\right ),\\
&A_6^{[2]}=A_6^{[5]}=\frac{1}{[1|5+6|4\rangle}\times\frac{1}{[3|1+2|6\rangle }\times\frac{S_{123}^3}{[12][23]\braket{45}\braket{56}}.
\end{aligned} \right .
\end{equation}
The results from the different decomposition show that the six-point amplitude \eqref{amplitude} can be reconstructed from the composition of inverse soft limit, since the central core of the soft limit of one particle turns out to be the soft shell the others. Our  six-point amplitude can be reconstructed from the combination of two inverse soft limits of two different five-points
\begin{equation}
 A_6(1^-,2^-,3^-,4^+,5^+,6^+)= S(6^+) A_5(1^-,2^-,3^-,4^+,5^+)+ S(1^-) A_5(2^-,3^-,4^+,5^+,6^+).
 \end{equation} 
This reconstruction of the amplitude \eqref{amplitude} can be represented in a diagrammatically as shown in Figure \ref{diagram_6}. The  investigation of the such reconstruction have been made for different size of amplitudes and for different helicity configurations. We found that the soft decomposition of MHV amplitudes,  helicity configurations with only two negative helicities \cite{Parke:1986gb}, doesn't have a soft shell which means they can be directly constructed from one single inverse soft limit of a lower point MHV. For higher point amplitudes, say $n$-points, the investigation of the different soft decomposition allows us to understand that the amplitude can be reconstructed from $(n-1)$-point amplitudes with at most ``three" linear combination of inverse soft limits. 

\begin{figure}
\centering
\begin{tikzpicture}[scale=.85,>=stealth',shorten >=1pt,node distance=2.5cm,on grid,initial/.style={rectangle},thick]
\def\p{*3};
\tikzset{mystyle/.style={->,very thick}}
\tikzset{every node/.style={fill=white}}
\node[state,initial] (T0) {$A_7(1^-,2^-,3^+,4^-,5^+,6^+,7^+)$};
\node[state,initial] (T3) at (0, 1\p) {$A_6(1^-,2^-,4^-,5^+,6^+,7^+)$};
\node[state,initial] (T34) at (1\p, 2\p) {$A_5(1^-,2^-,5^+,6^+,7^+)$};
\node[state,initial] (T35) at (-1\p, 2\p) {$A_5(1^-,2^-,4^-,6^+,7^+)$};
\node[state,initial](T2) at (1\p,-1\p) {$A_6(1^-,3^+,4^-,5^+,6^+,7^+)$};
\node[state,initial](T4) at (-1\p,-1\p) {$A_6(1^-,2^-,3^+,5^+,6^+,7^+)$};
\path (T3) edge [mystyle] node[pos=0.45] { $S(3^+)$} (T0);
\path (T34) edge [mystyle] node[pos=0.45] { $S(4^-)$} (T3);
\path (T35) edge [mystyle] node[pos=0.45] { $S({5}^+)$} (T3);
\path (T2) edge [mystyle] node[pos=0.45] { $S({2}^-)$} (T0);
\path (T4) edge [mystyle] node[pos=0.45] { $S({4}^-)$} (T0);
\end{tikzpicture}

\caption{Representation of the soft reconstruction of $A_7(1^-,2^-,3^+,4^-,5^+,6^+,7^+)$ }
\label{amplitude_7}\end{figure}
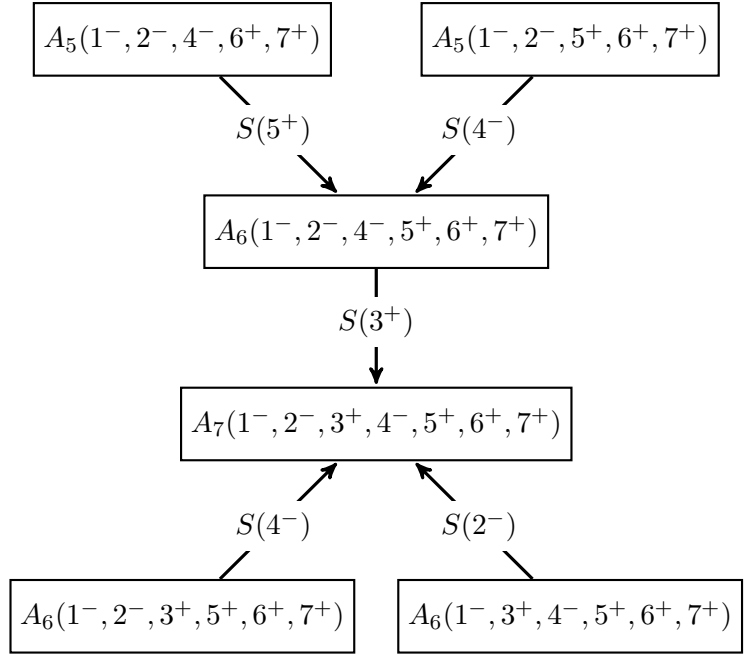

In Figure \ref{amplitude_7}, we can see that the typical 7-point amplitude with the helicity configuration $(1^-,2^-,3^+,4^-,5^+,6^+,7^+)$ can be reconstructed from the combination of three inverse soft limits of $A_6(1^-,2^-,4^-,5^+,6^+,7^+)$, $A_6(1^-,3^+,4^-,5^+,6^+,7^+)$, and $A_6(1^-,2^-,3^+,5^+,6^+,7^+)$. And in the same diagram we can also see that the $A_6(1^-,2^-,4^-,5^+,6^+,7^+)$ can be reconstructed from the combination of two inverse soft limits of $A_5(1^-,2^-,5^+,6^+,7^+)$ and $A_5(1^-,2^-,4^-,6^+,7^+)$.


\section{Conclusion}
In this work we would like to review the possibility to compute scattering amplitudes from them asymptotic symmetries. For that purpose we investigate the soft structure of amplitude, we called soft decomposition. We choose the six-point amplitude \eqref{amplitude} as our toy subject, in which we took the soft limit of particle and invert the limit using the inverse soft limit process. Depending on the helicity of the soft particle, the inverse limit consist to multiply the lower-point amplitude with the respective helicity soft factor, and to shift the two neighbor momenta of the soft inverted particle.

The process of the taking a particle soft then inverting the limit shows that amplitude can be decomposed as a combination of the central core of the soft limit, which is can be reconstructed directly from inverse soft, and a soft shell a piece that tend to zero as the particle goes soft \eqref{soft_decomposition}. Therefore, to construct an amplitude from inverse soft limit we need to find a way to compute the soft shell in the decomposition. In the investigation of the different decomposition of the amplitude, we also found that the soft shell of a given sodt decomposition is a linear combination of the different soft core of the different soft decomposition as show in Figure \ref{diagram_6} and Figure \ref{amplitude_7}.
From these results we found the importance of the soft cores of the different soft decomposition of a given amplitude in its soft construction, an amplitudes is a linear combination  of its soft cores. Our next step is now to find the right algorithm that combine the different soft decomposition in the inverse soft limit construction of amplitudes.

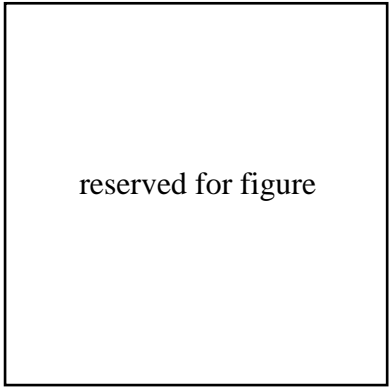
\end{document}